\documentclass[aps,pra,twocolumn,groupedaddress,showpacs,amsmath,amssymb,amsfonts,floatfix]{revtex4}

\usepackage{graphicx}
\usepackage{bm}

\renewcommand\labelenumi{(\roman{enumi})}
\renewcommand\theenumi\labelenumi

\begin{document}

\title{Two supersolid phases in hard-core extended Bose-Hubbard model}

\author{Yung-Chung Chen}
\affiliation{Department of Applied Physics, Tunghai University, Taichung 40704, Taiwan}

\author{Min-Fong Yang}
\affiliation{Department of Applied Physics, Tunghai University, Taichung 40704, Taiwan}

\date{\today}

\begin{abstract}
The effect of the next-nearest-neighbor (nnn) tunneling on the hard-core extended Bose-Hubbard model on square lattices is investigated. By means of the cluster mean-field theory, the ground-state phase diagrams are determined. When a modest nnn tunneling is introduced, depending on its sign, two distinct supersolid states with checkerboard crystal structures are found away from half-filing. The characters of various phase transitions out of these two supersolid states are discussed. In particular, for the case with kinetic frustration, the existence of a half supersolid phase possessing both solid and unconventional superfluid orders is established. Our work hence sheds light on the search of this interesting supersolid phase in real ultracold lattice gases with frustrated tunnelings.
\end{abstract}


\maketitle

\section{Introduction}

Ultracold atomic gases in optical lattices provide an extremely versatile and well-controlled experimental platform for exploring complex phenomena in quantum matters~\cite{BEC_review1,BEC_review2,BEC_review3}. Owing to remarkable control over microscopic model parameters, they can simulate the physics of strongly correlated systems in regimes which are not easily accessible in condensed matters. The rapid advances in experimental techniques open the door to the search for exotic phases and in probing their quantum critical behaviors.
One of the most striking experiments is the observation of the superfluid-Mott transition of ultracold bosons in an optical lattice~\cite{SF_Mott exp}, which can be described by the seminal Bose-Hubbard model~\cite{Fisher_etal_1989,Jaksch_etal_1998}. This quantum phase transition is driven by the competition between the kinetic energy of particles hopping between lattice sites and the on-site repulsive interaction.

Recently, by using the lattice shaking technique, the effective hopping parameters can be tuned to be negative or even complex~\cite{BEC_review3,%
Eckardt_etal_2005,Lignier_etal_2007,Struck_etal_2012,Eckardt_2017}. As a result, lattice bosons with frustrated hoppings and/or coupled to artificial gauge fields have now been realized, which can give rise to unconventional superfluid (SF) phases~\cite{Struck_etal_2011,Struck_etal_2013}.
Besides, intriguing supersolid (SS) states have been predicted to occur as well in the Bose-Hubbard model with hopping frustration induced by staggered flux~\cite{Moller_Cooper_2010}.

The SS phases are fascinating quantum states that have superfluid and solid long-range orders simultaneously. Such phases usually come out around the phase boundary of commensurate solid phases and behave as intermediate states between the superfluid and the crystalline ones. In addition to hopping frustration induced by the gauge fields~\cite{Moller_Cooper_2010}, ultracold lattice gases with dipole-dipole interaction~\cite{Lahaye_etal_2009,Baranov_etal_2012,Dutta_etal_2015} or the spin-orbit coupling~\cite{Li_etal_2013,Han_etal_2015,Li_etal_2017} also provide  possible candidates in simulating such quantum phases. In particular, the existence of the SS phases has been established in extended Bose-Hubbard models (i.e., Bose-Hubbard models with additional intersite interactions)~\cite{vanOtterlo_etal_1995,Goral_etal_2002,Sengupta_etal_2005,%
Yi_etal_2007,Danshita_SadeMelo_2009,Capogrosso-Sansone_etal_2010}, which can
be naturally realized with dipolar bosons. Due to intersite repulsions among lattice bosons, insulating solid states at commensurate filling factors can be stabilized. Under doping particles or holes into these insulating solid states, the SS states arises when dopants delocalize and undergo Bose-Einstein
condensation rather than a phase separation. The late realization of the extended Bose-Hubbard model with nearest-neighbour repulsion for magnetic erbium atoms lays the groundwork in searching for the SS states~\cite{Baier_etal_2016}.
Very recently, the SS phases are observed experimentally in atomic quantum gases trapped in an optical lattice inside a high-finesse optical cavity~\cite{Klinder_etal_2015,Landig_etal_2016}. They are stabilized by the long-range interaction mediated by a vacuum mode of the cavity.

More exotic states of matter are expected when one incorporates the hopping frustration into dipolar lattice gases. To uncover the interesting physics of such systems, an extended Bose-Hubbard model with a frustrated next-nearest-neighbor tunneling on square lattices is investigated recently~\cite{Dong_etal_2016}. In the hard-core limit, the grand-canonical Hamiltonian reads
\begin{eqnarray}\label{model}
H &=& -t \sum_{\langle i,j\rangle}(b_i^{\dag}b_j + h.c.) %
- t^\prime \sum_{\langle\langle i,j\rangle\rangle} (b_i^{\dag}b_j + h.c.) \nonumber \\
&&+V\sum_{\langle i,j\rangle} n_in_j - \mu\sum_i n_i \; ,
\end{eqnarray}
where $\langle i,j\rangle$ indicates the nearest- and $\langle\langle i,j\rangle\rangle$ next-nearest neighbors. The destruction (creation) operators of hard-core bosons on site $i$ are denoted by $b_i$ ($b_i^\dag$). The parameters $t$ and $t^\prime$ stand for the nearest-neighbor (nn) and the next-nearest-neighbor (nnn) hopping integrals, respectively. The repulsion between the nn sites is represented by $V$, $n_i=b_i^\dag b_i$ is the number operator of bosons, and $\mu$ is the chemical potential. Here both $t$ and $V>0$ are taken to be positive, while $t^\prime$ can be either positive or negative. When $t^\prime<0$, the hopping frustration appears.

In Ref.~\cite{Dong_etal_2016} where the frustrated case with $t^\prime<0$ is considered, by using the tensor network state method~\cite{TNS-1,TNS-2,TNS-3}, a peculiar half supersolid (HSS) phase is found to appear away from half-filling for large $V$ and modest $|t^\prime|$. Its stabilization results from the interplay of the frustrated nnn hopping and the nn repulsion. This HSS state is characterized by the preferential superfluid flow along the nnn bonds and nonzero Bose-Einstein condensate mainly on one sublattice. In addition, the particle density shows a checkerboard long-range order. The lattice's translational symmetry is thus spontaneously broken. A phase separation regime consisting of the uniform SF and the HSS phases is observed in their fixed-density models~\cite{Dong_etal_2016}. This implies a first-order phase transition between these two phases.

Actually, the non-frustrated nnn hopping ($t'>0$) can support as well a similar supersolid phase possessing both the SF and the checkerboard crystalline long-range orders~\cite{Chen_etal_2008}. The appearance of such a checkerboard supersolid (CSS) state for the postive-$t'$ case has been demonstrated by means of quantum Monte Carlo (QMC) simulations. In contrast to the the negative-$t'$ case, the SF-CSS phase transitions are however found to be of second-order~\cite{Chen_etal_2008}, instead of being discontinuous.

\begin{figure}
\includegraphics[clip,width=3.4in]{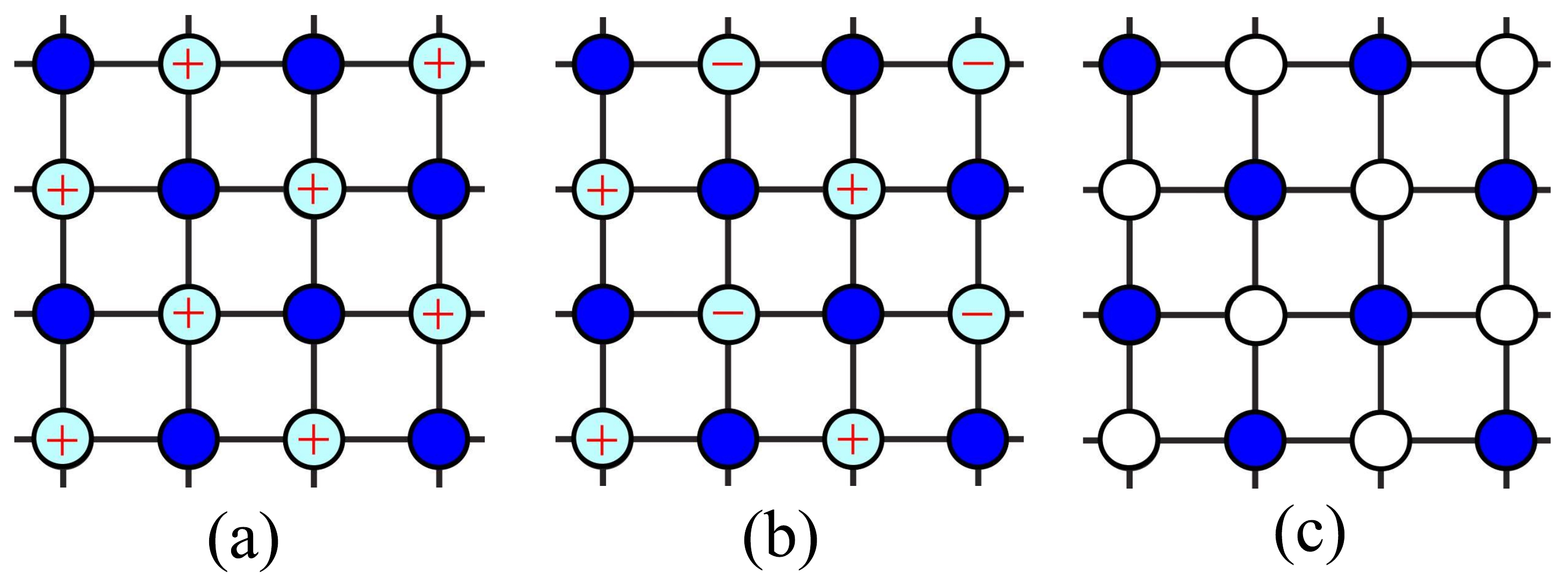}
\caption{Schematic pictures of (a) the checkerboard supersolid (CSS), (b) the half supersolid (HSS), and (c) the checkerboard solid (CBS). The empty (dark blue) circles indicate the sites of no (full or nearly full) occupation and the light blue circles stand for partially filled sites. The signs in the circles show the relative phases of Bose condensate among sites.
} \label{fig:phases}
\end{figure}

\begin{figure}
\includegraphics[trim={2.2cm 0.4cm 2cm 1.5cm},clip,width=3.54in]{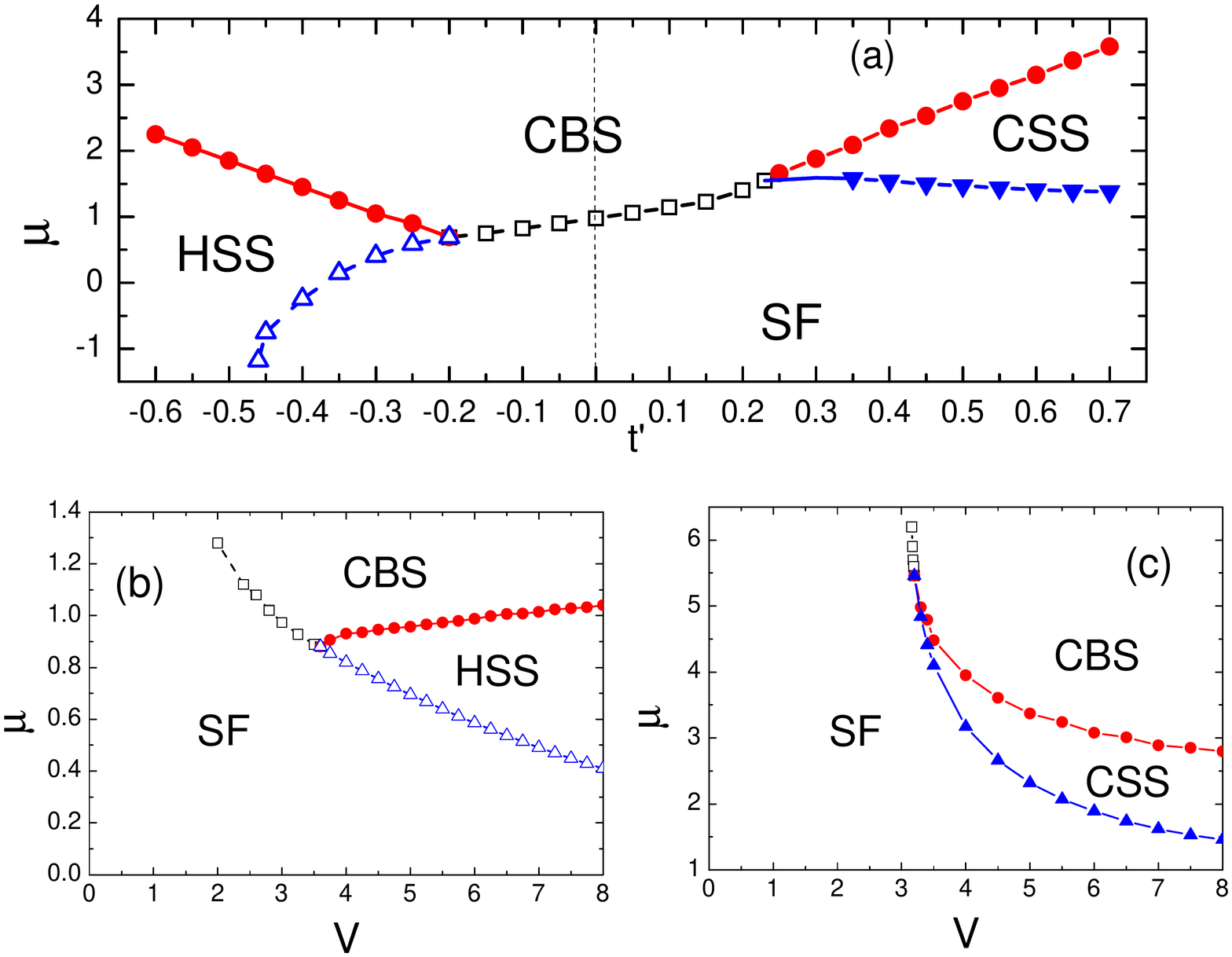}
\caption{Phase diagram calculated by the cluster mean-field theory using $4\times 4$ clusters for the extended Bose-Hubbard model in Eq.~\eqref{model} with fixed (a) $V=8$, (b) $t'=-0.3$, and (c) $t'=0.5$. Here $t=1$ is assumed. The solid and the open symbols indicate the second-order and the first-order transitions, respectively. The dashed (solid) lines connecting open (solid) symbols are only a guide to the eye.
The relative errors of the determined phase boundaries are set at the order of $10^{-2}$ and thus the error bars (not shown) for all the data points are smaller than the sizes of the symbols.
} \label{fig:phase_dia}
\end{figure}

In this paper, we focus on the detailed ground-state phase diagrams of the model in Eq.~\eqref{model} for both signs of $t'$ and elucidate the nature of the phase transitions therein. The cluster mean-field theory is employed here, which has been applied to various spin~\cite{Gelfand_etal_1989,Isaev_etal_2009-1,Isaev_etal_2009-2,yamamoto-14,%
Moreno-Cardoner_etal_2014} and boson~\cite{Hassan_etal_2007,yamamoto-12-1,yamamoto-12-2,McIntosh_etal_2012,%
Luhmann_2013,Huerga_etal_2013,Huerga_etal_2014,Singh_etal_2014,Jurgensen_etal_2015} systems with success.
We note that the main difference between two supersolid states, the HSS and the CSS states, comes from the distinct momentum states at which bosons condenses. Depending on the sign of $t'$, Bose condensate in the HSS phase occurs at the state with momenta $\textbf{k}=(0,\pi)$ and $(\pi,0)$, but at $\textbf{k}=(\pi,\pi)$ in the CSS phase. Schematic pictures of these two phases are illustrated in Figs.~\ref{fig:phases}(a) and (b). By measuring the types of condensation, both supersolid states can thus be identified.
Besides the uniform SF and the supersolid phases, there exists as well an incompressible checkerboard solid (CBS) state at half-filling [see Fig.~\ref{fig:phases}(c)]. This solid state breaks as well the lattice's translational symmetry and is signaled by a plateau in the density profile versus chemical potential.
Various transitions between these quantum phases are summarized in Fig.~\ref{fig:phase_dia}, which is the main result of this work.

For weak nnn hopping $t^\prime$, direct first-order transitions between the CBS and the SF phases are observed. Upon increasing $|t^\prime|$, an intermediate supersolid state (either CSS or HSS) emerges. In accordance with previous results~\cite{Dong_etal_2016,Chen_etal_2008}, we find that the SF-HSS transitions are of first order, while the SF-CSS transitions are continuous.
In contrast, both transitions from each supersolid phase to the CBS phase are shown to be continuous and belong to the superfluid-insulator universality class~\cite{Fisher_etal_1989}. According to our CMF results, the stability of the two supersolid phases is enhanced when the nn repulsion $V$ and the magnitude of the nnn hopping $t'$ are increased.
Since the model under consideration lies within the reach of present experimental setup, hopefully our predictions may be realized in the near future.

This paper is organized as follows: In Sec.~II, the cluster mean-field theory is briefly described and the measured order parameters are introduced. Our numerical results are presented and discussed in Sec.~III. We summarize our work in Sec.~IV.

\section{Cluster mean-field theory}

Our calculations are performed by using the cluster mean-field (CMF) theory~\cite{Gelfand_etal_1989,Isaev_etal_2009-1,Isaev_etal_2009-2,yamamoto-14,%
Moreno-Cardoner_etal_2014,Hassan_etal_2007,yamamoto-12-1,yamamoto-12-2,McIntosh_etal_2012,%
Luhmann_2013,Huerga_etal_2013,Huerga_etal_2014,Singh_etal_2014,Jurgensen_etal_2015}. It has several different names in the literature: cluster Gutzwiller approach~\cite{Luhmann_2013,Jurgensen_etal_2015}, hierarchical mean-field approach~\cite{Isaev_etal_2009-1,Isaev_etal_2009-2,Huerga_etal_2014}, and composite boson mean-field approach~\cite{Huerga_etal_2013}. This method has been shown to be an extremely efficient way of exploring vast uncharted territory in the phase diagram. Because short-range correlations within the cluster are taken into account exactly, the CMF method can give far more precise results than conventional single-site mean-field approaches. Under a scaling of the cluster sizes, even the accurate phase boundaries for infinite lattices may be achieved~\cite{yamamoto-12-1,yamamoto-12-2,Luhmann_2013}.

In the first subsection, we explain how the ground states are obtained under the CMF approach. In the second subsection, the ways in determining the order parameters and thus the ground-state phase diagrams are described. The method of the cluster-size scaling is discussed at the end.

\subsection{ground states in the CMF theory}

\begin{figure}
\includegraphics[clip,width=1.54in]{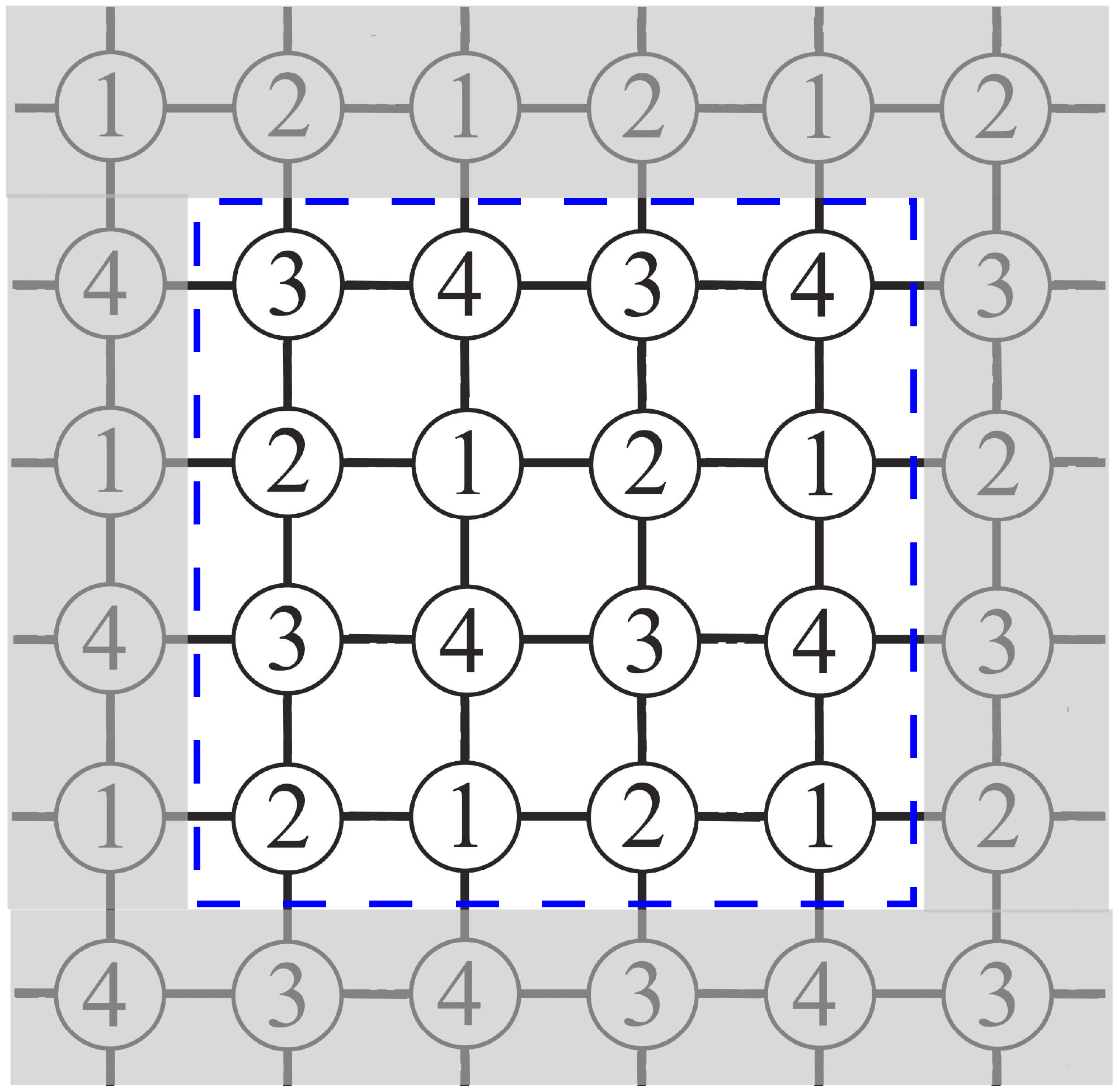}
\caption{A cluster of $4\times 4$ sites embedded into a four-sublattice structure. The numbers label the sublattices.
} \label{fig:CMF}
\end{figure}

In the CMF theory, the whole system is decomposed into a cluster of lattice sites and its surroundings, as illustrated in Fig.~\ref{fig:CMF}. The interaction between these two parts is approximated by the mean fields at the borders of the cluster. The effective Hamiltonian of the cluster is then given by
\begin{equation}\label{eff_Hami}
H_C^\textrm{eff} = H_C + H_{\partial C} \; ,
\end{equation}
where $H_C$ is the exact Hamiltonian of the cluster $C$ and $H_{\partial C}$ contains the effects of the mean fields acting only on sites at the boundary $\partial C$. By using the standard mean-field decoupling, the latter takes the form for our model in Eq.~\eqref{model},
\begin{eqnarray}\label{boundary_Hami}
H_{\partial C} &=& -t \sideset{}{'}\sum_{\langle i,j\rangle}
\left(b_i^{\dag} \langle b_j\rangle + h.c.\right)
-t^\prime \sideset{}{'}\sum_{\langle\langle i,j\rangle\rangle}
\left(b_i^{\dag} \langle b_j\rangle + h.c.\right) \nonumber \\
&& +V \sideset{}{'}\sum_{\langle i,j\rangle} n_i \langle n_j\rangle  \; ,
\end{eqnarray}
where the prime over $\sum$ indicates that the site $i$ locates on the boundary $\partial C$ and the site $j$ is outside the cluster. Such mean-field decoupling amounts to the approximation of the many-body wave function being written as a product of the cluster and the environment wave functions.

After diagonalizing the effective Hamiltonian $H_C^\textrm{eff}$, the mean fields of the surroundings, $\langle b_j\rangle$ and $\langle n_j\rangle$, can be obtained self-consistently from the corresponding expectation values inside the cluster. We note that the strategy in determining the values of mean fields is not unique. The implementation suggested in Refs.~\cite{yamamoto-12-1} and~\cite{yamamoto-12-2} is adopted here. A sublattice structure expected to emerge in the parameter range is first imposed, into which the cluster of $N_C$ sites is embedded. To be compatible with all the expected phases in our system (cf. Fig.~\ref{fig:phases}), states with a four-sublattice structure are considered in the present work. Assuming the mean fields $\langle b_j\rangle$ to be real, there are thus eight mean-field parameters, $\langle b_j\rangle$ and $\langle n_j\rangle$, in total. Under this four-sublattice pattern, a cluster of $N_C=4\times 4$ sites is illustrated in Fig.~\ref{fig:CMF}. The mean fields to which the cluster is coupled are then calculated by averaging over all sites belonging to the same sublattice. In some cases, this prescription can prevent possible numerical instability happened in other proposals~\cite{Moreno-Cardoner_etal_2014}.

Given the system parameters $t$, $t^\prime$, $V$, and $\mu$, the self-consistent procedure for determining the ground state includes the following key steps:
\begin{enumerate}
\setlength{\itemsep}{-3pt}
  \item\label{item:1} Substitute a set of initial mean-field values of $\langle b_j\rangle$ and $\langle n_j\rangle$ into Eq.~\eqref{boundary_Hami}, and calculate the ground state $|\textrm{GS}'\rangle$ of the effective Hamiltonian $H_C^\textrm{eff}$ in Eq.~\eqref{eff_Hami}.
  \item\label{item:2} Calculate the new mean-field values $\langle b_j'\rangle=\langle\textrm{GS}'|b_j|\textrm{GS}'\rangle$ and $\langle n_j'\rangle=\langle\textrm{GS}'|n_j|\textrm{GS}'\rangle$ for the ground state $|\textrm{GS}'\rangle$.
  \item\label{item:3} Compare two sets of mean fields, \{$\langle b_j\rangle$, $\langle n_j\rangle$\} and \{$\langle b_j'\rangle$, $\langle n_j'\rangle$\}. If the net deviation is smaller than the given tolerance, that is, $\sum_j |\langle b_j'\rangle-\langle b_j\rangle| + \sum_j |\langle n_j'\rangle-\langle n_j\rangle| < \epsilon$ (in the present work, we take $\epsilon=10^{-6}$), then output $|\textrm{GS}'\rangle$ as the ground state, $\langle b_j'\rangle$ and $\langle n_j'\rangle$ as the self-consistent mean fields. Otherwise, set $\langle b_j\rangle$=$\langle b_j'\rangle$, $\langle n_j\rangle$=$\langle n_j'\rangle$ and return to step~\ref{item:1}.
  \end{enumerate}

\subsection{order parameters and phase diagram}

To characterize different phases, several order parameters are calculated.
Nonzero Bose condensate indicates the presence of the SF long-range order. The condensate density $\rho_0$ at a momentum $\textbf{k}$ can be evaluated through the Fourier transform of the mean fields $\langle b_j\rangle$,
\begin{equation}\label{SF}
\rho_0(\textbf{k})\equiv \left|\frac{1}{N_C} \sum_{j\in C} \left\langle b_j  \right\rangle e^{i\textbf{k}\cdot\textbf{r}_j}\right|^2 \; ,
\end{equation}
where $N_C$ is the total number of sites in the cluster $C$. In a uniform SF phase, we have Bose condensate at $\textbf{k}=(0,0)$ with $\rho_0(0,0)\neq0$ and $\rho_0(\textbf{k})=0$ for all other $\textbf{k}$'s. Instead, unconventional SF orders are found in the CSS and the HSS phases. From the inspection of their condensation patterns depicted in Fig.~\ref{fig:phases}, one expects that both $\rho_0(0,0)$ and $\rho_0(\pi,\pi)$ are nonzero in the CSS phase, while only $\rho_0(\pi,0)=\rho_0(0,\pi)$ can be finite in the HSS phase.

On the other hand, long-range solid order appears if there exists density modulation with a commensurate wave vector $\textbf{k}$. This can be detected by nonzero values of the Fourier transform of the local densities $\langle n_j \rangle$,
\begin{equation}\label{den_mod}
\tilde{n}(\textbf{k}) \equiv \frac{1}{N_C} \sum_{j\in C} \left\langle n_j  \right\rangle e^{i\textbf{k}\cdot\textbf{r}_j} \; .
\end{equation}
Its value at $\textbf{k}=(0,0)$ is nothing but the averaged density. In the thermodynamic limit, this quantity has a direct relation, $|\tilde{n}(\textbf{k})|^2 =S(\textbf{k})/N_C$, to the static structure factor $S(\textbf{k}) \equiv (1/N_C)\sum_{i,j\in C} \left\langle n_i n_j \right\rangle e^{i\textbf{k}\cdot(\textbf{r}_i-\textbf{r}_j)}$. For systems with the CBS order, $\tilde{n}(\pi,\pi)\neq0$ is expected. In particular, the classical picture of the incompressible CBS state at half-filling predicts $\tilde{n}(\pi,\pi)=1/2$.

Besides, in order to determine possible first-order phase transitions signaled by energy level crossing~\cite{Sachdev:book},
we need to measure the ground-state energies of our systems. In the CMF theory, the expression of the energy per site is
\begin{equation}\label{energy}
e = \frac{1}{N_C}\left(E_G - \frac{1}{2} \langle H_{\partial C}\rangle \right) \; ,
\end{equation}
where $E_G$ is the lowest energy eigenvalue of the effective Hamiltonian $H_C^\textrm{eff}$. We note that double counting terms coming from the mean-field decoupling of two-cluster interactions must be subtracted. This explains the occurrence of the second term in Eq.~\eqref{energy}.

The phase diagrams presented in Fig.~\ref{fig:phase_dia} are obtained by means of the CMF approach using $4\times4$ clusters. We note that the phase boundaries will be slightly affected upon increasing the cluster sizes, since longer-ranged many-particle correlations can be incorporated. By applying a cluster-size scaling, the phase transition points for an infinite lattice can be extrapolated~\cite{yamamoto-12-1,yamamoto-12-2,Luhmann_2013}. In this work, we take the scaling parameter $\lambda=N_B/(N_C\times z/2)$ introduced in Refs.~\cite{yamamoto-12-1} and~\cite{yamamoto-12-2}. Here, $N_B$ denotes the number of the nn bonds within the cluster and $z$ is the coordination number for the lattice ($z=4$ for the present case of square lattice). Because the denominator represents the number of the nn bonds of the original lattice per $N_C$ sites, the parameter $\lambda$ provides an indication of how much the correlation effects between particles are considered in the cluster. The infinite-size limit thus corresponds to $\lambda=1$.
In the next section, linear extrapolations toward $\lambda=1$ for some typical phase transition points are performed to illustrate the cluster-size dependence of our results.

\section{Numerical results and discussions}

\subsection{non-frustrated case of $t'>0$}

\begin{figure}
\includegraphics[trim={3cm 0.9cm 2.5cm 1cm},clip,width=3.4in]{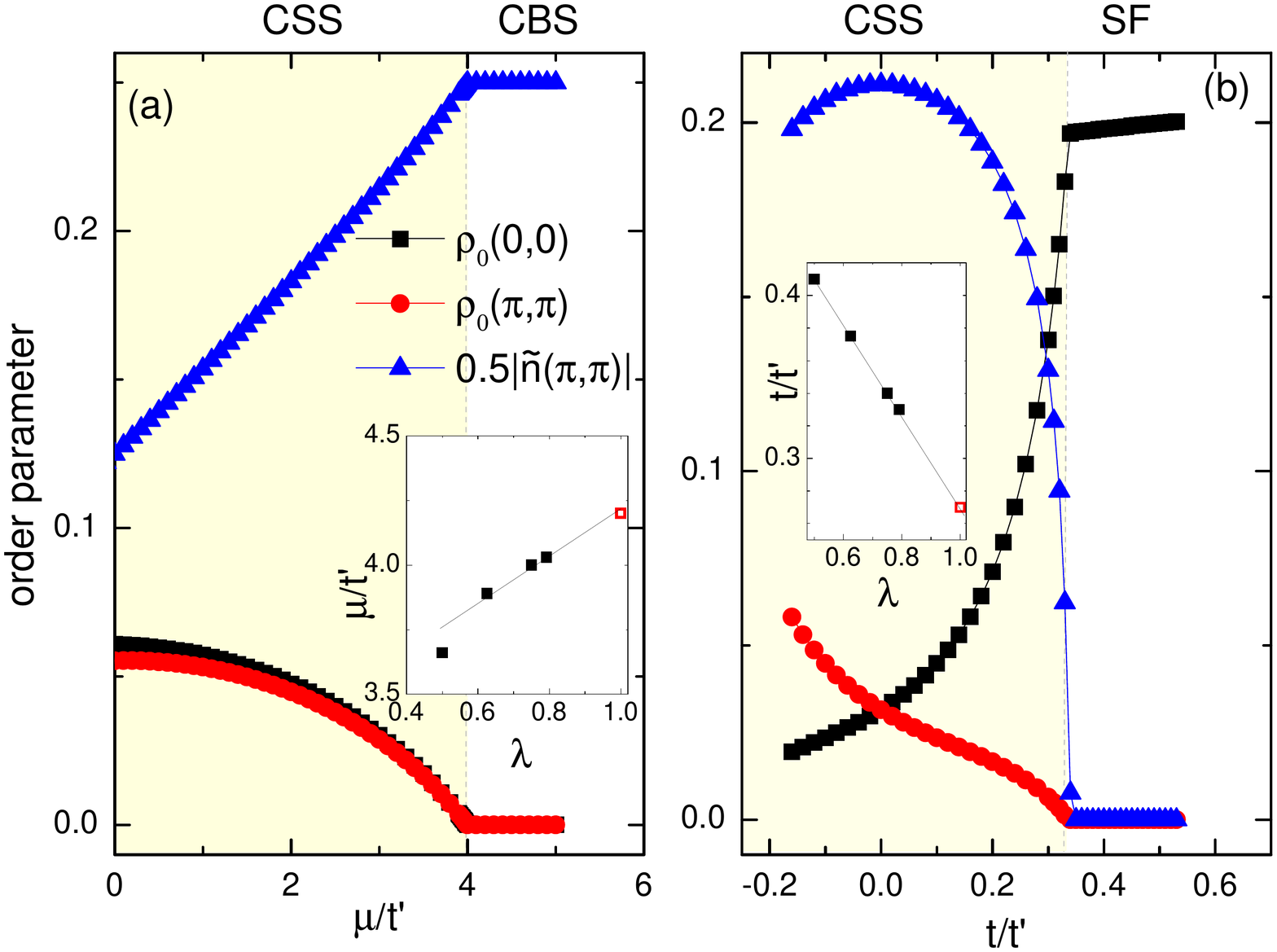}
\caption{Dependence of various order parameters (a) on the chemical potential $\mu$ for $V=3.5\,t'$ and $t=0.01\,t'$; (b) on the nn hopping $t$ for $V=3.2\,t'$ and $\mu=2.9\,t'$, which are calculated by using $4\times 4$ clusters. Here $t'=1$ is taken as energy unit. In both cases, $\rho_0(\pi,0)=\rho_0(0,\pi)=0$.
Insets in the panels (a) and (b) show the cluster-size scalings of the CMF data for the phase transition points $\mu_c$ and $t_c$, respectively. The scaling parameter $\lambda=0.5$, 0.625, 0.75, 0.79 correspond respectively to the clusters of rectangular geometry with sizes $N_C=2\times2$, $2\times4$, $4\times4$, and $4\times6$. The lines are the linear fits of the values for the largest three sizes. The red open symbols indicate the QMC results~\cite{Chen_etal_2008}.}
\label{fig:benchmark}
\end{figure}

In order to provide benchmarks on the performance of the CMF method, we begin with the non-frustrated case of $t'>0$, where some QMC data are available in the literature~\cite{Chen_etal_2008}. Fig.~\ref{fig:benchmark}(a) shows our CMF results of various order parameters as functions of the chemical potential $\mu$ for $V/t'=3.5$ and $t/t'=0.01$ obtained by using $4\times 4$ clusters. For this special case of $t\ll t'$, upon decreasing $\mu$, the incompressible CBS state with $\tilde{n}(\pi,\pi)=1/2$ and $\rho_0(0,0)=\rho_0(\pi,\pi)=0$ transits continuously into the CSS state with coexistence of the CBS and the SF orders. This continuous melting transition occurs at $\mu_c/t'\simeq4.0$. Our findings are in accordance with those calculated by the QMC approach (see Fig.~3 in Ref.~\cite{Chen_etal_2008}), in which the critical value $\mu_c/t'\simeq4.2$ is observed. As seen from the inset, the values of this transition point will be shifted to larger ones when the cluster size is increased. Excellent agreement can be reached after a linear extrapolation toward the scaling parameter $\lambda=1$. As a further benchmark, the CMF results using $4\times 4$ clusters for $V/t'=3.2$ and $\mu/t'=2.9$ are presented in Fig.~\ref{fig:benchmark}(b). By increasing the nn hopping $t$, the uniform SF state will be eventually stabilized, such that $\tilde{n}(\pi,\pi)$ vanishes and $\rho_0(\textbf{k})$ is nonzero for $\textbf{k}=(0,0)$ only. The CSS-SF transition is found to be as well continuous and it occurs at $t_c/t'\simeq0.34$. They are consistent with the QMC results (see the lowest panel of Fig.~4 in Ref.~\cite{Chen_etal_2008}), where $t_c/t'\simeq0.27$ is found. Again, excellent agreement can be obtained after a linear extrapolation toward the scaling parameter $\lambda=1$, as shown in the inset of Fig.~\ref{fig:benchmark}(b). All these results validate the CMF approach to the present hard-core boson problem.

The requirement of $t<t'$ studied in Ref.~\cite{Chen_etal_2008} is hard to be satisfied in most cold atom experiments. One may wonder if similar physics remains in the more practical situation such that the nn hopping is larger than the nnn one (i.e., $t>t'$). In the rest of this paper, only weak $t'$ cases are considered, and we take $t=1$ as the energy unit.

\begin{figure}
\includegraphics[trim={3cm 0.3cm 2.5cm 0.3cm},clip,width=3.4in]{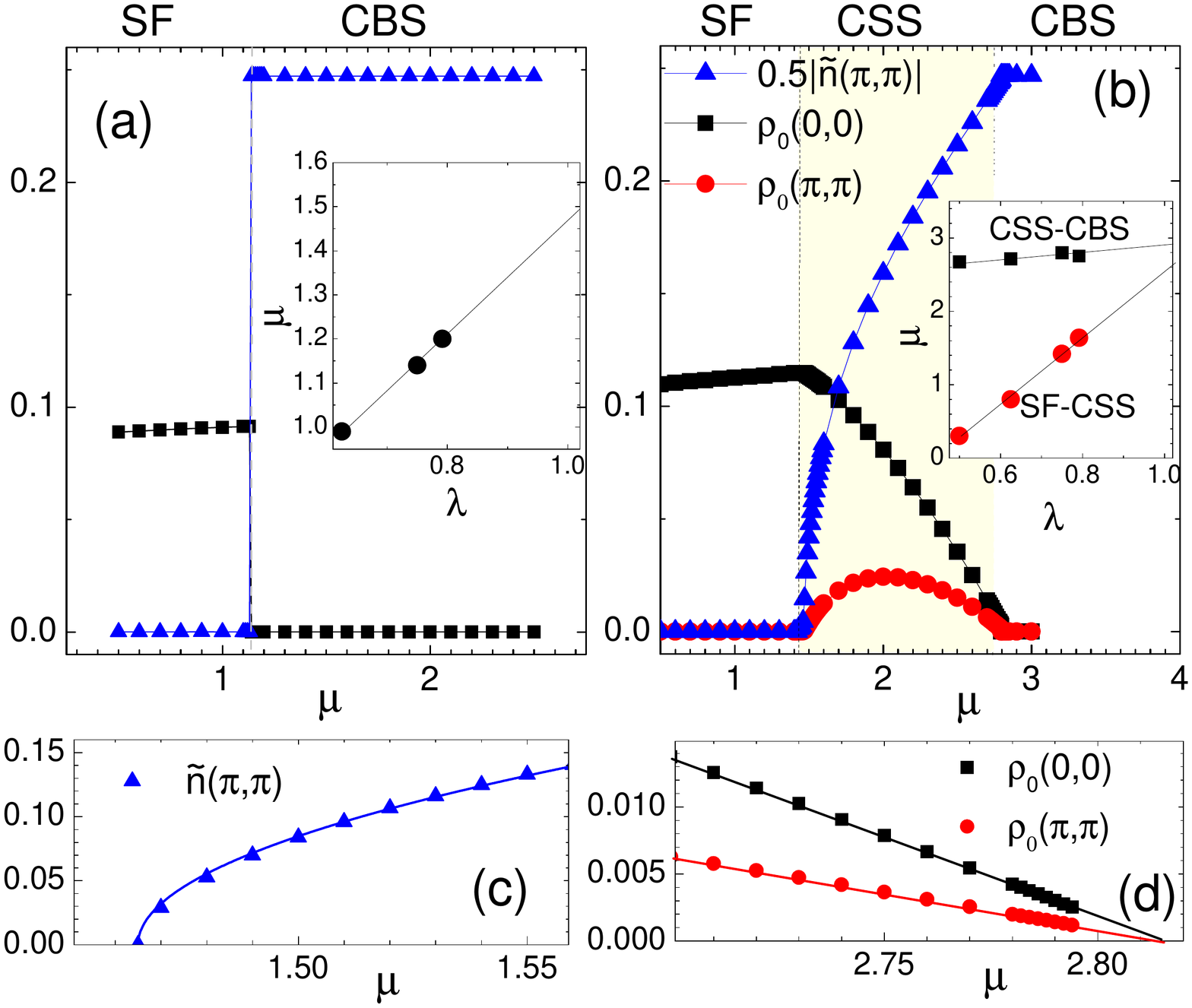}
\caption{Various order parameters as functions of the chemical potential $\mu$ calculated by using $4\times 4$ clusters with fixed $V=8$ for (a) $t'=0.1$ and (b) $t'=0.5$. Here $t=1$ is taken as energy unit. In both cases, $\rho_0(\pi,0)=\rho_0(0,\pi)=0$.
Insets in the panels show the cluster-size scalings of the CMF data for the phase transition points $\mu_c$. The scaling procedure is discussed in Fig.~\ref{fig:benchmark}.
Lower panels show the details of the order parameters around the SF-CSS and the CSS-CBS transition points in panel (b) with enlarged scales. Lines in these panels are (c) the square-root and (d) the linear fits of the data.
}
\label{fig:positive_tp}
\end{figure}

In Fig.~\ref{fig:positive_tp}, the CMF results using $4\times 4$ clusters for two typical values of $t'=0.1$ and 0.5 with fixed $V=8$ are presented. For the $t'=0.1$ case shown in panel (a), a direct SF-CBS transition is observed, at which the SF and the CBS order parameters, $\rho_0(0,0)$ and $\tilde{n}(\pi,\pi)$, change discontinuously. This first-order transition occurs at $\mu_c\simeq1.14$. As seen from the inset, the transition point approaches to $\mu_c\simeq1.49$ in the infinite-size limit (i.e., $\lambda\to 1$). The absence of the supersolid phase for this weak $t'$ case is expected. In the $t'=0$ limit, it has been known that the hard-core boson model~\cite{Batrouni_Scalettar_1995,Batrouni_Scalettar_2000,Hebert_etal_2001} (or the equivalent spin-1/2 $XXZ$ model~\cite{Kohno_Takahashi_1997,Yunoki_2002}) does not stabilize the supersolid phase on a square lattice and the direct SF-CBS transition is of first order. Turning on a small $t'$ would not change the whole story, as shown by our results.

According to the Ginzburg-Landau-Wilson paradigm, a direct transition between the SF and the CBS states, which break different symmetries, must be of first order. Otherwise, an intermediate supersolid phase with coexistence of SF and solid long-range orders will emerge in between. The occurrence of the CSS phase between the SF and the CBS phases at modest nnn hoppings has been established in Fig.~\ref{fig:positive_tp}(b). For $t'=0.5$, within our CMF results using $4\times 4$ clusters,  the CSS phase is stabilized when $1.46\lesssim\mu\lesssim2.80$. The transitions at these two phase boundaries are both continuous. As seen from the inset, the stability region of the CSS phase shrinks but remains finite in the infinite-size limit (i.e., $\lambda\to 1$).

Two transitions out of the CSS phase have distinct characters. The SF-CSS transition is expected to be of Ising type, since the two-fold sublattice symmetry broken in the CSS state due to the CBS order will be restored after this transition. Around this critical point, our findings of the CBS order parameter $\tilde{n}(\pi,\pi)$ show a square-root dependence on $\mu$, $\tilde{n}(\pi,\pi)\propto(\mu-\mu_c)^{1/2}$, as presented in Fig.~\ref{fig:positive_tp}(c). That is, the critical exponent has a mean-field value $\beta=1/2$. This reflects the mean-field character of our approach. On the other hand, the CSS-CBS transition should belong to the SF-insulator universality class~\cite{Fisher_etal_1989}, such that the condensate fraction vanishes linearly around the transition point. Our results of both condensate fractions $\rho_0(0,0)$ and $\rho_0(\pi,\pi)$ shown in Fig.~\ref{fig:positive_tp}(d) are in agreement with this prediction.

The phase diagrams for the extended Bose-Hubbard model in Eq.~\eqref{model} with $t'>0$ is depicted in Figs.~\ref{fig:phase_dia}(a) and (c). We find that the CSS phase exists in extended regions of the phase diagrams. From our CMF calculations, besides the $t'\gg t$ limit investigated in Ref.~\cite{Chen_etal_2008}, this phase can appear as well for weak $t'$ cases as long as strong nn repulsion $V$ is present.

\subsection{frustrated case of $t'<0$}

Now we turn to the frustrated case of $t'<0$. Due to the negative-sign problem caused by the hopping frustration, QMC simulations are no longer applicable. Therefore, to obtain quantitative predictions of the present frustrated systems, other theoretical approaches, such as the tensor network state method employed in Ref.~\cite{Dong_etal_2016}, are necessary. With success in the $t'>0$ case presented above, we continue our CMF studies on the $t'<0$ side.

\begin{figure}
\includegraphics[trim={2.4cm 0.3cm 1.8cm 0.3cm},clip,width=3.4in]{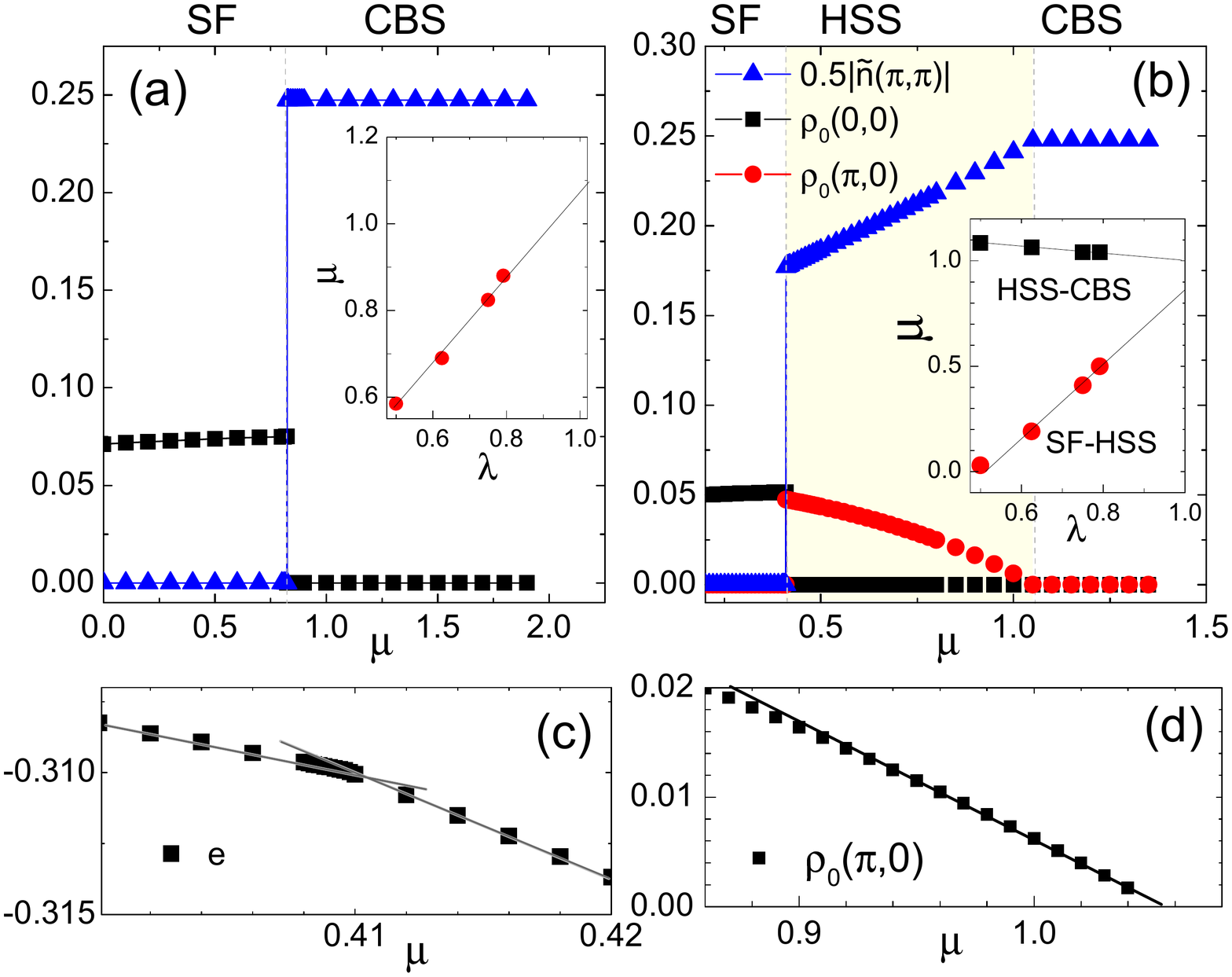}
\caption{Various order parameters as functions of the chemical potential $\mu$ calculated by using $4\times 4$ clusters with fixed $V=8$ for (a) $t'=-0.1$ and (b) $t'=-0.3$. Here $t=1$ is taken as energy unit. In both cases, $\rho_0(\pi,\pi)=0$ and $\rho_0(0,\pi)=\rho_0(\pi,0)$.
Insets in the panels show the cluster-size scalings of the CMF data for the phase transition points $\mu_c$. The scaling procedure is discussed in Fig.~\ref{fig:benchmark}.
Panel (c) shows the ground-state energies $e$ as a function of $\mu$ calculated by Eq.~\eqref{energy} around the SF-HSS transition point in panel (b). Two straight gray lines connecting symbols are only a guide to the eye. They display the level crossing and the cusp in energy curve. Panel (d) gives the condensate fraction $\rho_0(\pi,0)$ and its linear fit around the HSS-CBS transition point in panel (b) with an enlarged scale.
}
\label{fig:negative_tp}
\end{figure}

Our CMF results using $4\times 4$ clusters for two typical values of $t'=-0.1$ and $-0.3$ with fixed $V=8$ are presented in Fig.~\ref{fig:negative_tp}. Similar to the cases of weak positive $t'$, a direct first-order transition between the SF and the CBS phases is observed for the $t'=-0.1$ case [see Fig.~\ref{fig:negative_tp}(a)]. This SF-CBS transition occurs at $\mu_c\simeq0.82$ and the value of the transition point approaches to $\mu_c\simeq1.09$ in the infinite-size limit (i.e., $\lambda\to 1$).

Upon increasing the magnitude of the negative $t'$, an intermediate HSS phase with coexistence of SF and solid long-range orders can emerge in between. Similar to the CSS phase in the non-frustrated case, the HSS state is characterized as well by the preferential superfluid flow along the nnn bonds. Nevertheless, Bose condensate in the HSS phase occurs at the state with momenta $\textbf{k}=(0,\pi)$ and $(\pi,0)$, instead of $\textbf{k}=(\pi,\pi)$ in the CSS phase. The case of $t'=-0.3$ is shown in Fig.~\ref{fig:negative_tp}(b), in which the HSS phase is stabilized when $0.41\lesssim\mu\lesssim1.05$ within our CMF results using $4\times 4$ clusters. As seen from the inset, the stability region of the HSS phase shrinks but remains finite in the infinite-size limit (i.e., $\lambda\to 1$).

In agreement with the SF-insulator universality class~\cite{Fisher_etal_1989}, we find that the condensate fraction $\rho_0(\pi,0)$ vanishes linearly around the continuous HSS-CBS transition [see Fig.~\ref{fig:negative_tp}(d)]. However, the phase transition between the uniform SF and the present supersolid phases is found to be of first order, instead of being continuous in the non-frustrated case of $t'>0$. The discontinuous jumps in the order parameters at the SF-HSS transition point is a clear indication that phase separation would occur in a canonical system, as observed in Ref.~\cite{Dong_etal_2016}.

First-order quantum phase transitions come from energy level crossing in the ground state and are thus signaled by a cusp in the ground-state energy curve~\cite{Sachdev:book}. In Fig.~\ref{fig:negative_tp}(c), the energy curve in the ground state around the discontinuous SF-HSS transition is shown, where the location of the cusp gives the transition point we want. In the present work, all the first-order transition points, including those of the direct SF-CBS transitions shown in Figs.~\ref{fig:positive_tp}(a) and~\ref{fig:negative_tp}(a), are determined by this way.

Distinct characters of the SF-CSS transition for the $t'>0$ case and the SF-HSS transition for the $t'<0$ case can be understood from different condensation patterns in these two supersolid states. Bose condensate at the state with momenta $\textbf{k}=(\pi,0)$ and $(0,\pi)$ in the HSS state gives %
$\left\langle b_j\right\rangle \propto e^{i(\pi,0)\cdot\textbf{r}_j} + e^{i(0,\pi)\cdot\textbf{r}_j} = \cos(\pi x) + \cos(\pi y)$. %
This leads to alternating phase change in $\pi$ among sites within the sublattice on which bosons condense, as displayed in Fig.~\ref{fig:phases}(b). Thus a smooth evolution from the HSS state to the uniform SF state should be unlikely. On the contrary, there is no phase difference in the condensate for the CSS state [see Fig.~\ref{fig:phases}(a)]. A continuous transition from the CSS state to the uniform SF state is thus expected.

The phase diagrams for the extended Bose-Hubbard model in Eq.~\eqref{model} with $t'<0$ is presented in Figs.~\ref{fig:phase_dia}(a) and (b). The existence of the HSS phase in extended regions of the phase diagrams is demonstrated. Indicated by our CMF calculations, the stability of the HSS phase is enhanced when the nn repulsion $V$ and the magnitude of the nnn hopping $t'$ are increased. As seen from the phase diagrams in Fig.~\ref{fig:phase_dia}, the HSS phase for $t'<0$ exists in larger parameter regions than the CSS phase for $t'>0$ does. This implies that, in stabilizing the supersolid phase, the effect of frustrated nnn hopping is more significant. In particular, when $t'/t<-1/2$, the uniform SF phase becomes absent and the HSS phase can exist for all incommensurate fillings. This can be understood from the single-particle dispersion relation, which has local minima at $\mathbf{k}=(0,\pi)$ and $\mathbf{k}=(\pi,0)$ when $t'/t<-1/2$. This thus favors the Bose condensation consistent with the SF order in the HSS state, rather than the uniform SF one.

\section{Summary and conclusions}

Using the cluster mean-field theory, we explored in detail the formation of two supersolid phases, which arise in the hard-core extended boson Hubbard model of Eq.~\eqref{model}. We find that, in the presence of a large nn repulsion, by introducing a modest nnn hopping $t'$, either the CSS or the HSS phases can be stabilized away from half-filling. Both supersolid phases come out around the phase boundary of the CBS phase and behave as intermediate states between the SF and the CBS states. Despite carrying the same checkerboard crystalline long-range order, bosons condense at different momentum states in these two supersolid phases, as shown in our calculations.

The characters of various phase transitions out of the CSS or the HSS states are discussed. Their transitions to the CBS state are found to be continuous and belong to the SF-insulator universality class~\cite{Fisher_etal_1989}. However, for the frustrated case with $t'<0$, the SF-HSS transition is shown to be of first order, instead of being continuous observed at the SF-CSS transition for the $t'>0$ case. The distinct nature of these two phase transitions can be understood from their different condensation patterns.

While only the model of hard-core bosons is considered in this work, similar phenomena are expected as well when the hard-core constraint is replaced by a finite but strong on-site repulsion. The latter condition can be accomplished, for example, by tuning the s-wave scattering length via a Feshbach resonance~\cite{Inouye_etal_1998,Chin_etal_2010}. As mentioned in the Introduction, the realization of the extended Bose-Hubbard model has been achieved recently~\cite{Baier_etal_2016} and frustrated tunnelings can be produced by, say, lattice shaking techniques~\cite{BEC_review3,%
Eckardt_etal_2005,Lignier_etal_2007,Struck_etal_2012,Eckardt_2017}. That is, systems related to the model discussed here will be hopefully soon available experimentally. Our results thus provide a useful guide to the experimental search of these supersolid phases, especially the interesting HSS phase in real ultracold lattice gases with frustrated tunnelings.

\begin{acknowledgments}
Y.C.C. and M.F.Y. acknowledge the support from the Ministry of Science and
Technology of Taiwan under Grant No. NSC 102-2112-M-029-003-MY3, MOST 105-2112-M-029-005, MOST 106-2112-M-029-003, and MOST 106-2112-M-029-004. We also thank Kwai-Kong Ng for the critical reading and useful discussions.
\end{acknowledgments}


\end{document}